\documentclass[10pt,journal,compsoc]{IEEEtran}

\usepackage[noadjust]{cite}
\usepackage{filecontents}

\usepackage{graphics}
\usepackage{lipsum}
\usepackage{graphicx,amsmath,alltt}
\usepackage[linesnumbered,ruled]{algorithm2e}
\usepackage{textcomp}
\usepackage{booktabs}
\usepackage{multirow}
\usepackage{colortbl}
\usepackage{ragged2e}
\usepackage{tabularx,booktabs}
\usepackage[dvipsnames,table,xcdraw]{xcolor}
\renewcommand{\citedash}{--}

\SetKwComment{Comment}{/* }{ */}

\begin{document}


\title{
\normalfont{Technical Report} \\
\normalfont\Large\bfseries{Solving the Job Shop Scheduling Problem with Ant Colony Optimization}
}

\author{
    \IEEEauthorblockN {
       Alysson~Ribeiro~da~Silva
    }
   
    \IEEEauthorblockA {
        Computer Science Graduate Program \\
        Federal University of Minas Gerais \\
        Belo Horizonte, Brazil
    }
}

\IEEEtitleabstractindextext{%


\justify
\begin{abstract}
The Job Shop Schedule Problem (JSSP) refers to the ability of an agent to allocate tasks that should be executed in a specified time in a machine from a cluster. The task allocation can be achieved from several methods, however, this report it is explored the ability of the Ant Colony Optimization to generate feasible solutions for several JSSP instances. This proposal models the JSSP as a complete graph since disjunct models can prevent the ACO from exploring all the search space. Several instances of the JSSP were used to evaluate the proposal. Results suggest that the algorithm can reach optimum solutions for easy and harder instances with a selection of parameters.
\end{abstract}


}

\maketitle
\IEEEdisplaynontitleabstractindextext
\IEEEpeerreviewmaketitle
\ifCLASSOPTIONcompsoc


\section{Introduction}

Ant Colony Optimization is an optimization method from Evolutionary Computing that was inspired by ant colony behavior \cite{Blum2004, Puris2007, Turguner2014, Zhang2010}. It allows for optimization processes through traces of pheromones left by simulated ants through the best-found path towards a valid solution in a graph. For example, several ants explore the search space, in a directed graph, and when reaching a terminal vertex, the solution is extracted from the found path. At the end of the process, the simulated ants deposit pheromones that are utilized by other ants to reach the same solution or to explore the space near it. In this report, the ACO was used to solve the Job Shop Schedule Problem (JSSP).

The JSSP can be typically modeled as a disjunctive graph, however, due to its nature in this report, the problem was modeled as a complete graph. The distance between each graph edge was calculated as a delta Makespan, where the objective is to minimize the total Makespan. Furthermore, in this report, a solution is obtained from a Hamiltonian path followed by each ant, where the global Makespan is generated with an efficient algorithm.

All the routines were implemented in C++ due to performance issues. The experiments were conducted in two steps. The first step comprises the parameter evaluation, whereas in the second the algorithm's best solutions were evaluated. Four instances of the JSSP from \cite{Instances2020} were used for evaluation, the \textbf{ft06}, \textbf{la01}, \textbf{la29}, and \textbf{la49}. The parameter evaluation was conducted with \textbf{ft06} and \textbf{la29}. Differently, the algorithms' performance was evaluated in all instances. The proposal shows convergence for all instances towards the global optimum. At the end of this report is shown the evaluation of solutions obtained from the ACO for several other JSSP instances.

\section{Job Shop Scheduling Problem}

The JSSP can be defined as an optimization problem, where the objective is to minimize the total Makespan. Let $J = \{j_1,...,j_n\}$ be the set of all services, $M = \{m_1,...,m_m\}$ the set of machines, $O_{ij} = \{O_{i1},...,O_{{ik}_i}\}$ the set of operations $J_i$ that need to be executed in the machine $M_k$ for $\tau_{ik}$, e $\tau_{ik} = \{\tau_{i1},...,\tau_{{ik}_i}\}$ time period, the objective is to:

\begin{equation}
    \begin{split}
        \text{Minimize } C_{max} & = max(t_{ik} + \tau_{ik}), \text{$\forall J_i \in J$ e $M_k \in M$} \\
        \tau_{ik} & \ge 0 \\
        t_{ik} - t_{ih} & \ge \tau_{ih} \text{ if $O_{ih}$ precedes $O_{ik}$} \\
        t_{pk} - t_{ik} + K(1 - \gamma_{ipk}) & \ge \tau_{ik} \\
        t_{ik} - t_{pk} + K(\gamma_{ipk}) & \ge \tau_{pk} \\
        \gamma_{ipk} & = 1 \text{ if $O_{ik}$ precedes $O_{pk}$} \\
        \gamma_{ipk} & = 0 \text{ otherwise}
    \end{split}
\end{equation}

\noindent
where $i, p \in J$ e $k, h \in M$ and $K > \sum_i (\sum_k \tau_{ik} - min(\tau_{ik}))$.
\section{Method}

This proposal uses the ACO to find the best possible Makespan. The problem was modeled as a complete graph similar to a disjunctive graph approach \cite{Puris2007}. Next, the ACO is presented alongside the JSSP model.

\subsection{Elitist ACO}

For the optimization of the Makespan, it is proposed to use an elitist ACO, where the transition between vertices is done through the Equation~\ref{eq:path_prob},

\begin{equation}
    \label{eq:path_prob}
    p_{ij} = \frac{(\tau_{ij})^\alpha(\eta_{ij}^\beta)} {\sum_l (\tau_{il})^\alpha (\eta_{il})^\beta}
\end{equation}

\noindent
where, $p_{ij}$ is the transition probability from $i$ to $j$, $\tau_{ij}$ is the pheromone amount for arc $i \rightarrow j$, $\tau_{ij}$ represents the distance between $i$ and $j$, and $\sum_l (\tau_{il})^\alpha (\eta_{il})^\beta$ is the sum of the pheromones times the distance for all available transitions from vertex $i$.

The pheromones decay is computed with Equation~\ref{eq:decay}, where, $d$ represents the decaying factor for the pheromone quantity $\tau_{ij}$.

\begin{equation}
    \label{eq:decay}
    \tau_{ij} = (1 - d) \tau_{ij}
\end{equation}

Pheromones update is achieved through Equation~\ref{eq:reinforce}, where $L_k$ is the total Makespan generated by the Ant $k$ if $k$ passed through the arc $i \rightarrow j$.

\begin{equation}
    \label{eq:reinforce}
    \tau_{ij} = \tau_{ij} + \sum_k \frac{Q}{L_k}
\end{equation}

\subsection{JSSP Model}

The graph model used to represent tasks and resources is inspired by disjunctive graphs. A disjunctive graph is defined as $G = (V, C \cup D)$, where $V$ is the set of all tasks, $C$ is the edges that connect tasks from the same service, and $D$ represents edges that connect tasks from the same machine. The problem with this representation regards the fact that it does not allow an Ant to generate all possible feasible solutions. Consequently, in this research the search graph is defined as $G' = (V, C \cup D \cup F)$, where $F$ contains the complement of $G$. To avoid invalid paths, each Ant should check for two restrictions in this model. The first prevents from selecting tasks with parents yet running and the second prevents from choosing tasks already chosen.

\subsection{Distance Between Tasks}

A distance between tasks is computed as $1/d$ \cite{Blum2004, Puris2007, Turguner2014, Zhang2010} to minimize the total Makespan, where $d$ is the time necessary to complete a task. However, this computation can lead to distances that do not represent the real impact of the chosen task in reality. Consequently, it is proposed that a distance $\eta_{ij}$ between $i$ and $j$ to be computed as $\Delta m = s - c$, where $m$ is the current Makespan from a partial path to $i$, and $s$ is the future possible Makespan if $j$ is chosen. To extract a solution from the generated path, a Gantt diagram is built by following the best-found policy in $G'$.
\section{Experimental Configuration}

Two-stage experimentation was conducted on an AMD CPU using Ubuntu 20.04. In the first stage, the impact of the parameters of the ACO was evaluated. Differently, the second stage evaluates the solutions obtained by the algorithm when solving the JSSP. The ACO was deployed in C++ with OpenMP due to performance concerns. 

Four instances of the JSSP were primarily used for the experiments and they are described in Table \ref{tab:inst}. For the parameter evaluation, only the \texttt{ft06} and \texttt{la01} were used since they portray traits of all four instances. Finally, the quality of the generated solution was evaluated for the best-observed parameters for all instances of Table \ref{tab:inst}. At the end of the document, it is provided the evaluation of the algorithm for other instances.

\begin{table}[h]
\caption{Instances for the JSSP problem.}
\begin{align*}
\begin{tabular}{lcc}
\textbf{Inst.} & \multicolumn{1}{l}{\textbf{Jobs x Machines}} & \multicolumn{1}{l}{\textbf{Optimum}} \\ \cline{2-3} 
\multicolumn{1}{l|}{\textbf{ft06}} & \multicolumn{1}{c|}{6x6} & \multicolumn{1}{c|}{55} \\ \cline{2-3} 
\multicolumn{1}{l|}{\textbf{la01}} & \multicolumn{1}{c|}{10x5} & \multicolumn{1}{c|}{666} \\ \cline{2-3} 
\multicolumn{1}{l|}{\textbf{la29}} & \multicolumn{1}{c|}{20x10} & \multicolumn{1}{c|}{1157} \\ \cline{2-3} 
\multicolumn{1}{l|}{\textbf{la40}} & \multicolumn{1}{c|}{15x15} & \multicolumn{1}{c|}{1222} \\ \cline{2-3} 
\end{tabular}
\end{align*}
\label{tab:inst}
\end{table}

All results convey the statistics from $30$ execution with $1000$ iterations each. Each execution was done in parallel with OpenMp. The number of iterations was fixed at $1000$ to avoid large memory consumption when running in parallel. The initial and minimum pheromone value for all evaluations was set to $1$ to avoid normalization issues in case $\eta_{ij} < 1$.

\section{Parameter Analysis}
\label{sec:param}

All parameters were evaluated according to the configurations in Table~\ref{tab:param}. Each cell in the table that is depicted with the orange color represents the main parameters. The columns \textit{Elit.}, \textit{Init.}, and \textit{Inc.} represent flags that can be active or not the elitism, random initialization, and pheromones deposit policy, respectively. In special, the \textit{Init.} flag can assume three values that define how the initial vertex is chosen for each Ant, such as:

\begin{itemize}
    \item 0) Random.
    \item 1) Random at the first iteration and fixed at the others.
    \item 2) Only one Ant for service. If the operation mode is set to 2, the Ants quantity is ignored and only one Ant is created for each service.
\end{itemize}

Differently, the \textit{Inc.} flag defines which policy will be used to deposit pheromones. If set to $0$, then the Equation~\ref{eq:reinforce} is used. Otherwise, Equation~\ref{eq:reinforce_ex} is used,

\begin{equation}
    \label{eq:reinforce_ex}
    \tau_{ij} = \tau_{ij} + (\frac{Q}{L_k})^{|P| - i}
\end{equation}

\noindent
where $P$ represents the path generated by the Ant $K$, $|P|$ is the total path length, and $i$ represents a vertex from $P$. Consequently, it is possible to ensure that vertices closest to the object will receive more pheromones.

\begin{table*}[ht]
\caption{Configurations to check parameters influence in the ACO's performance.}
\begin{align*}
\begin{tabular}{ccccccccccc}
\textbf{Config.}                      & \textbf{Ite.}                                     & \textbf{Ants.}           & \textbf{Elit.}                                 & \textbf{Alpha}                                 & \textbf{Beta}                                  & \textbf{Evap.}                                    & \textbf{Q}             & \textbf{Init}                                  & \textbf{Exec.}          & \textbf{Inc.}                                  \\ \cline{2-11} 
\multicolumn{1}{c|}{\textbf{elitism}} & \multicolumn{1}{c|}{1000}                         & \multicolumn{1}{c|}{100} & \multicolumn{1}{c|}{\cellcolor[HTML]{FFCE93}1} & \multicolumn{1}{c|}{1}                         & \multicolumn{1}{c|}{1}                         & \multicolumn{1}{c|}{0.1}                          & \multicolumn{1}{c|}{1} & \multicolumn{1}{c|}{0}                         & \multicolumn{1}{c|}{30} & \multicolumn{1}{c|}{0}                         \\ \cline{2-11} 
\multicolumn{1}{c|}{\textbf{alpha}}   & \multicolumn{1}{c|}{1000}                         & \multicolumn{1}{c|}{100} & \multicolumn{1}{c|}{0}                         & \multicolumn{1}{c|}{\cellcolor[HTML]{FFCE93}2} & \multicolumn{1}{c|}{1}                         & \multicolumn{1}{c|}{0.1}                          & \multicolumn{1}{c|}{1} & \multicolumn{1}{c|}{0}                         & \multicolumn{1}{c|}{30} & \multicolumn{1}{c|}{0}                         \\ \cline{2-11} 
\multicolumn{1}{c|}{\textbf{beta}}    & \multicolumn{1}{c|}{1000}                         & \multicolumn{1}{c|}{100} & \multicolumn{1}{c|}{0}                         & \multicolumn{1}{c|}{1}                         & \multicolumn{1}{c|}{\cellcolor[HTML]{FFCE93}2} & \multicolumn{1}{c|}{0.1}                          & \multicolumn{1}{c|}{1} & \multicolumn{1}{c|}{0}                         & \multicolumn{1}{c|}{30} & \multicolumn{1}{c|}{0}                         \\ \cline{2-11} 
\multicolumn{1}{c|}{\textbf{evap. A}} & \multicolumn{1}{c|}{1000}                         & \multicolumn{1}{c|}{100} & \multicolumn{1}{c|}{0}                         & \multicolumn{1}{c|}{1}                         & \multicolumn{1}{c|}{1}                         & \multicolumn{1}{c|}{\cellcolor[HTML]{FFCE93}1.0}  & \multicolumn{1}{c|}{1} & \multicolumn{1}{c|}{0}                         & \multicolumn{1}{c|}{30} & \multicolumn{1}{c|}{0}                         \\ \cline{2-11} 
\multicolumn{1}{c|}{\textbf{evap. B}} & \multicolumn{1}{c|}{1000}                         & \multicolumn{1}{c|}{100} & \multicolumn{1}{c|}{0}                         & \multicolumn{1}{c|}{1}                         & \multicolumn{1}{c|}{1}                         & \multicolumn{1}{c|}{\cellcolor[HTML]{FFCE93}0.5}  & \multicolumn{1}{c|}{1} & \multicolumn{1}{c|}{0}                         & \multicolumn{1}{c|}{30} & \multicolumn{1}{c|}{0}                         \\ \cline{2-11} 
\multicolumn{1}{c|}{\textbf{evap. C}} & \multicolumn{1}{c|}{1000}                         & \multicolumn{1}{c|}{100} & \multicolumn{1}{c|}{0}                         & \multicolumn{1}{c|}{1}                         & \multicolumn{1}{c|}{1}                         & \multicolumn{1}{c|}{\cellcolor[HTML]{FFCE93}0.01} & \multicolumn{1}{c|}{1} & \multicolumn{1}{c|}{0}                         & \multicolumn{1}{c|}{30} & \multicolumn{1}{c|}{0}                         \\ \cline{2-11} 
\multicolumn{1}{c|}{\textbf{init. A}} & \multicolumn{1}{c|}{1000}                         & \multicolumn{1}{c|}{100} & \multicolumn{1}{c|}{0}                         & \multicolumn{1}{c|}{1}                         & \multicolumn{1}{c|}{1}                         & \multicolumn{1}{c|}{0.1}                          & \multicolumn{1}{c|}{1} & \multicolumn{1}{c|}{\cellcolor[HTML]{FFCE93}0} & \multicolumn{1}{c|}{30} & \multicolumn{1}{c|}{0}                         \\ \cline{2-11} 
\multicolumn{1}{c|}{\textbf{init. B}} & \multicolumn{1}{c|}{1000}                         & \multicolumn{1}{c|}{100} & \multicolumn{1}{c|}{0}                         & \multicolumn{1}{c|}{1}                         & \multicolumn{1}{c|}{1}                         & \multicolumn{1}{c|}{0.1}                          & \multicolumn{1}{c|}{1} & \multicolumn{1}{c|}{\cellcolor[HTML]{FFCE93}1} & \multicolumn{1}{c|}{30} & \multicolumn{1}{c|}{0}                         \\ \cline{2-11} 
\multicolumn{1}{c|}{\textbf{init. C}} & \multicolumn{1}{c|}{1000}                         & \multicolumn{1}{c|}{100} & \multicolumn{1}{c|}{0}                         & \multicolumn{1}{c|}{1}                         & \multicolumn{1}{c|}{1}                         & \multicolumn{1}{c|}{0.1}                          & \multicolumn{1}{c|}{1} & \multicolumn{1}{c|}{\cellcolor[HTML]{FFCE93}2} & \multicolumn{1}{c|}{30} & \multicolumn{1}{c|}{0}                         \\ \cline{2-11} 
\multicolumn{1}{c|}{\textbf{ite. A}}  & \multicolumn{1}{c|}{\cellcolor[HTML]{FFCE93}100}  & \multicolumn{1}{c|}{100} & \multicolumn{1}{c|}{0}                         & \multicolumn{1}{c|}{1}                         & \multicolumn{1}{c|}{1}                         & \multicolumn{1}{c|}{0.1}                          & \multicolumn{1}{c|}{1} & \multicolumn{1}{c|}{0}                         & \multicolumn{1}{c|}{30} & \multicolumn{1}{c|}{0}                         \\ \cline{2-11} 
\multicolumn{1}{c|}{\textbf{ite B.}}  & \multicolumn{1}{c|}{\cellcolor[HTML]{FFCE93}500}  & \multicolumn{1}{c|}{100} & \multicolumn{1}{c|}{0}                         & \multicolumn{1}{c|}{1}                         & \multicolumn{1}{c|}{1}                         & \multicolumn{1}{c|}{0.1}                          & \multicolumn{1}{c|}{1} & \multicolumn{1}{c|}{0}                         & \multicolumn{1}{c|}{30} & \multicolumn{1}{c|}{0}                         \\ \cline{2-11} 
\multicolumn{1}{c|}{\textbf{ite C.}}  & \multicolumn{1}{c|}{\cellcolor[HTML]{FFCE93}1000} & \multicolumn{1}{c|}{100} & \multicolumn{1}{c|}{0}                         & \multicolumn{1}{c|}{1}                         & \multicolumn{1}{c|}{1}                         & \multicolumn{1}{c|}{0.1}                          & \multicolumn{1}{c|}{1} & \multicolumn{1}{c|}{0}                         & \multicolumn{1}{c|}{30} & \multicolumn{1}{c|}{0}                         \\ \cline{2-11} 
\multicolumn{1}{c|}{\textbf{Inc.}}    & \multicolumn{1}{c|}{1000}                         & \multicolumn{1}{c|}{100} & \multicolumn{1}{c|}{0}                         & \multicolumn{1}{c|}{1}                         & \multicolumn{1}{c|}{1}                         & \multicolumn{1}{c|}{0.1}                          & \multicolumn{1}{c|}{1} & \multicolumn{1}{c|}{0}                         & \multicolumn{1}{c|}{30} & \multicolumn{1}{c|}{\cellcolor[HTML]{FFCE93}1} \\ \cline{2-11} 
\end{tabular}
\end{align*}
\label{tab:param}
\end{table*}

The results for all the configuration parameter selection from Table~\ref{tab:param} is shown in Fig.~\ref{fig:ft06_param} for the instance \textit{ft06}. The $Y$ axis represents the average Makespace for $40$ executions and the $X$ axis shows each parameter configuration. According to the obtained results, the worst performance was $58.7^{+_0.68}_{-0.89}$ with a deviation of $1.28$. This behavior could indicate that in this particular instance the pheromones can lead Ants far from the optimum. Differently, the configurations \textit{Beta}, \textit{Evap. C}, and \textit{Inc.} were able to obtain near optimum Makespans. In particular, the configuration \textit{Evap. C}, which regards modifications in the pheromone's evaporation rate, obtained the optimum Makespan with $0$ deviation. The observed behavior could indicate that a smoother pheromones evaporation helps Ants to retain more memory regarding the environment exploration and consequently to explore new areas surrounding a solution. Other configurations show an average Makespan between $56$ and $58$ with subtle changes in deviation.

\begin{figure}[h]
    \centering
    \includegraphics[width=0.45\textwidth]{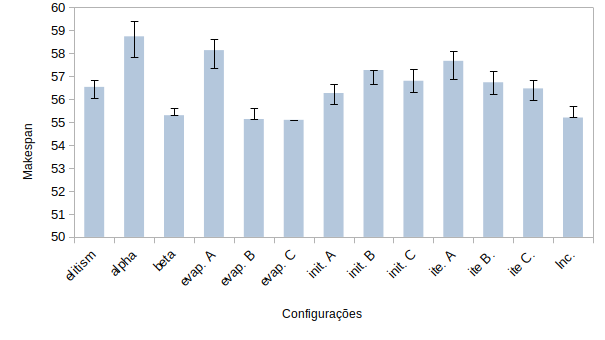}
    \caption{Average, Sup. Std., and Inf. Std. for all parameter configurations for instance ft06.}
    \label{fig:ft06_param}
\end{figure}

Fig.~\ref{fig:la01_param} shows the average Makespan for all parameters in the \textit{la01} instance. Configuration \textit{Init. C} obtained the worst results with an average Makespan of $676^{+4.99}_{-3.03}$ and a deviation of $8.5$. The aforementioned results could be explained as disturbances caused by the initial vertex selection mechanism. On the other hand, the best makespan was obtained with configuration \textit{Inc.} as a global optimum with $0$ deviation. These results corroborate with a low evaporation rate as happened for instance \textit{ft06}.

\begin{figure}[h]
    \centering
    \includegraphics[width=0.45\textwidth]{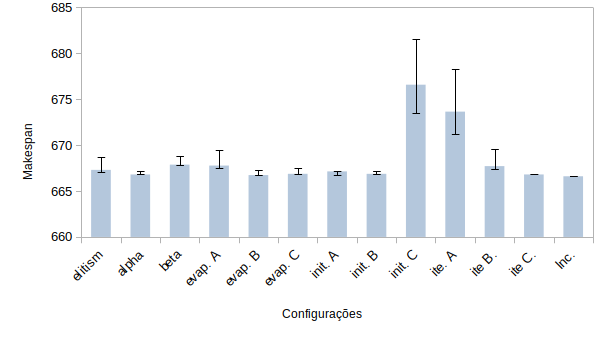}
    \caption{Average, Sup. Std., and Inf. Std. for all parameter configurations for instance la01.}
    \label{fig:la01_param}
\end{figure}

\section{Parameter Selection}

The selected parameters for evaluation are described in Table~\ref{tab:param_perf}, where the selection was made based on convergence, deviation, and best obtained Makespan.

\begin{table*}[ht]
\caption{Selected parameters for evaluation.}
\centering
\begin{tabular}{lllllllllll}
\textbf{Inst.} & \textbf{Ite.} & \textbf{Ants.} & \textbf{Elit.} & \textbf{Alpha} & \textbf{Beta} & \textbf{Evap.} & \textbf{Q} & \textbf{Init} & \textbf{Exec.} & \textbf{Inc.} \\ \cline{2-11} 
\multicolumn{1}{l|}{\textbf{ft06}} & \multicolumn{1}{c|}{1000} & \multicolumn{1}{c|}{-} & \multicolumn{1}{c|}{1} & \multicolumn{1}{c|}{1} & \multicolumn{1}{c|}{2} & \multicolumn{1}{c|}{0.01} & \multicolumn{1}{c|}{1} & \multicolumn{1}{c|}{2} & \multicolumn{1}{c|}{30} & \multicolumn{1}{c|}{1} \\ \cline{2-11} 
\end{tabular}
\label{tab:param_perf}
\end{table*}

\begin{table*}[t]
\caption{Obtained Makespan for all evaluation instances.}
\centering
\begin{tabular}{lccccccc}
\textbf{Inst.} & \multicolumn{1}{l}{\textbf{Optimum}} & \multicolumn{1}{l}{\textbf{Minimum}} & \multicolumn{1}{l}{\textbf{Maximum}} & \multicolumn{1}{l}{\textbf{Average}} & \multicolumn{1}{l}{\textbf{Std.}} & \multicolumn{1}{l}{\textbf{Sup. Std.}} & \multicolumn{1}{l}{\textbf{Inf. Std.}} \\ \cline{2-8} 
\multicolumn{1}{l|}{\textbf{ft06}} & \multicolumn{1}{c|}{55} & \multicolumn{1}{c|}{55} & \multicolumn{1}{c|}{58} & \multicolumn{1}{c|}{56,57} & \multicolumn{1}{c|}{1,09} & \multicolumn{1}{c|}{0,49} & \multicolumn{1}{c|}{0,5} \\ \cline{2-8} 
\multicolumn{1}{l|}{\textbf{la01}} & \multicolumn{1}{c|}{666} & \multicolumn{1}{c|}{666} & \multicolumn{1}{c|}{687} & \multicolumn{1}{c|}{673,07} & \multicolumn{1}{c|}{6,36} & \multicolumn{1}{c|}{3,99} & \multicolumn{1}{c|}{2,08} \\ \cline{2-8} 
\multicolumn{1}{l|}{\textbf{la29}} & \multicolumn{1}{c|}{1157} & \multicolumn{1}{c|}{1388} & \multicolumn{1}{c|}{1455} & \multicolumn{1}{c|}{1429,67} & \multicolumn{1}{c|}{17} & \multicolumn{1}{c|}{7,17} & \multicolumn{1}{c|}{13,69} \\ \cline{2-8} 
\multicolumn{1}{l|}{\textbf{la40}} & \multicolumn{1}{c|}{1222} & \multicolumn{1}{c|}{1333} & \multicolumn{1}{c|}{1381} & \multicolumn{1}{c|}{1361,67} & \multicolumn{1}{c|}{12,06} & \multicolumn{1}{c|}{4,66} & \multicolumn{1}{c|}{9,14} \\ \cline{2-8} 
\end{tabular}
\label{tab:main}
\end{table*}

The total iterations were set to $1000$ to avoid memory consumption and statistics were collected during $30$ executions. Evaporation was set to $0.1$ for slow convergence and stability. Next, elitism was used to ensure the best solution for newly developed generations. Differently, the Beta parameter was set to $2$ for a greater influence of the delta Makespan in the generated paths and to avoid pheromone degradation. Finally, $Q$ was set to $1$ for normalization purposes in the final solution.

\section{Results}

All the results of the algorithm's performance are shown in Table~\ref{tab:main}. It was observed that the algorithm was able to reach a global optimum in both, \textit{ft06} and \textit{la01} instances, instances. In \textit{ft06}, the average Makespace was $56.57^{+0.49}_{-0.5}$. Differently, \textit{la01} portrayed an average Makespace of $673.07^{+3.99}_{-2.08}$. Furthermore, the algorithm obtained sub-optimum average Makespans of $1429.67^{+7.17}_{-13.69}$ and $1361.67^{+4.66}_{-9.14}$ for instances \textit{la29} and \textit{la40}, respectively. The observed behavior could have happened due to several factors, for example, the found path, pheromones reinforce parameter, Ants path selection policy, evaporation rate, and other parameters. Also, the fitness landscape could also have influenced the performance of the algorithms, since the aforementioned instances are harder.
\section{Conclusion}

In this technical report, the capability of the ACO (Ant Colony Optimization) was evaluated to solve the JSSP (Job Shop Schedule Problem) problem. Several instances were evaluated and the algorithm was able to reach the optimum value for simple instances. Furthermore, it reached near-optimum ones in harder instances. Nevertheless, evaluation of several other instances of the JSSP presented in Appendix A, shows that it can solve the problem with low standard deviation in several scenarios. The parameter selection plays an important role in the algorithm's performance since for different parameters the algorithm was not able to achieve optimum results.


\bibliographystyle{IEEEtran}
\bibliography{main}

\clearpage

\begin{appendices}
\section{Other Instances Evaluation}
\label{appendix:A}

\begin{table}[t]
\centering
\caption{ACO best-found solution for extra instances.}
\label{tab:extraresults}
\begin{tabular}{llllllll}
\textbf{Inst.} & \textbf{Optimum} & \textbf{Minimum} & \textbf{Maximum} & \textbf{Average} & \textbf{Std.} & \textbf{Std. Sup.} & \textbf{Std Inf.} \\ \cline{2-8} 
\multicolumn{1}{l|}{\textbf{abz5}} & \multicolumn{1}{l|}{1234} & \multicolumn{1}{l|}{1272} & \multicolumn{1}{l|}{1303} & \multicolumn{1}{l|}{1289,4} & \multicolumn{1}{l|}{11,77} & \multicolumn{1}{l|}{0,50} & \multicolumn{1}{l|}{6,34} \\ \cline{2-8} 
\multicolumn{1}{l|}{\textbf{abz9}} & \multicolumn{1}{l|}{678} & \multicolumn{1}{l|}{810} & \multicolumn{1}{l|}{834} & \multicolumn{1}{l|}{821,4} & \multicolumn{1}{l|}{8,24} & \multicolumn{1}{l|}{4,00} & \multicolumn{1}{l|}{4,50} \\ \cline{2-8} 
\multicolumn{1}{l|}{\textbf{orb10}} & \multicolumn{1}{l|}{944} & \multicolumn{1}{l|}{1019} & \multicolumn{1}{l|}{1074} & \multicolumn{1}{l|}{1046,8} & \multicolumn{1}{l|}{18,80} & \multicolumn{1}{l|}{9,67} & \multicolumn{1}{l|}{7,50} \\ \cline{2-8} 
\multicolumn{1}{l|}{\textbf{swv05}} & \multicolumn{1}{l|}{1424} & \multicolumn{1}{l|}{1757} & \multicolumn{1}{l|}{1772} & \multicolumn{1}{l|}{1765,8} & \multicolumn{1}{l|}{5,81} & \multicolumn{1}{l|}{2,83} & \multicolumn{1}{l|}{2,50} \\ \cline{2-8} 
\multicolumn{1}{l|}{\textbf{swv19}} & \multicolumn{1}{l|}{2843} & \multicolumn{1}{l|}{3026} & \multicolumn{1}{l|}{3052} & \multicolumn{1}{l|}{3039,8} & \multicolumn{1}{l|}{9,81} & \multicolumn{1}{l|}{1,00} & \multicolumn{1}{l|}{4,50} \\ \cline{2-8} 
\multicolumn{1}{l|}{\textbf{swv20}} & \multicolumn{1}{l|}{2823} & \multicolumn{1}{l|}{2936} & \multicolumn{1}{l|}{2997} & \multicolumn{1}{l|}{2969,6} & \multicolumn{1}{l|}{20,07} & \multicolumn{1}{l|}{10,03} & \multicolumn{1}{l|}{13,50} \\ \cline{2-8} 
\multicolumn{1}{l|}{\textbf{yn1}} & \multicolumn{1}{l|}{884} & \multicolumn{1}{l|}{1022} & \multicolumn{1}{l|}{1046} & \multicolumn{1}{l|}{1035,6} & \multicolumn{1}{l|}{8,91} & \multicolumn{1}{l|}{3,30} & \multicolumn{1}{l|}{3,50} \\ \cline{2-8} 
\multicolumn{1}{l|}{\textbf{yn2}} & \multicolumn{1}{l|}{870} & \multicolumn{1}{l|}{1030} & \multicolumn{1}{l|}{1070} & \multicolumn{1}{l|}{1059,2} & \multicolumn{1}{l|}{14,85} & \multicolumn{1}{l|}{3,04} & \multicolumn{1}{l|}{0,00} \\ \cline{2-8} 
\multicolumn{1}{l|}{\textbf{yn3}} & \multicolumn{1}{l|}{859} & \multicolumn{1}{l|}{1008} & \multicolumn{1}{l|}{1038} & \multicolumn{1}{l|}{1026,2} & \multicolumn{1}{l|}{10,89} & \multicolumn{1}{l|}{2,87} & \multicolumn{1}{l|}{6,00} \\ \cline{2-8} 
\multicolumn{1}{l|}{\textbf{yn4}} & \multicolumn{1}{l|}{929} & \multicolumn{1}{l|}{1145} & \multicolumn{1}{l|}{1170} & \multicolumn{1}{l|}{1157,4} & \multicolumn{1}{l|}{8,31} & \multicolumn{1}{l|}{4,00} & \multicolumn{1}{l|}{4,71} \\ \cline{2-8} 
\multicolumn{1}{l|}{\textbf{dmu01}} & \multicolumn{1}{l|}{2501} & \multicolumn{1}{l|}{3097} & \multicolumn{1}{l|}{3193} & \multicolumn{1}{l|}{3150,6} & \multicolumn{1}{l|}{37,10} & \multicolumn{1}{l|}{0,50} & \multicolumn{1}{l|}{18,52} \\ \cline{2-8} 
\multicolumn{1}{l|}{\textbf{dmu20}} & \multicolumn{1}{l|}{3604} & \multicolumn{1}{l|}{4771} & \multicolumn{1}{l|}{4864} & \multicolumn{1}{l|}{4806,4} & \multicolumn{1}{l|}{36,64} & \multicolumn{1}{l|}{14,50} & \multicolumn{1}{l|}{5,73} \\ \cline{2-8} 
\multicolumn{1}{l|}{\textbf{dmu50}} & \multicolumn{1}{l|}{3496} & \multicolumn{1}{l|}{4648} & \multicolumn{1}{l|}{4715} & \multicolumn{1}{l|}{4689,2} & \multicolumn{1}{l|}{23,73} & \multicolumn{1}{l|}{9,74} & \multicolumn{1}{l|}{17,00} \\ \cline{2-8} 
\multicolumn{1}{l|}{\textbf{dmu80}} & \multicolumn{1}{l|}{6459} & \multicolumn{1}{l|}{8674} & \multicolumn{1}{l|}{8832} & \multicolumn{1}{l|}{8757,8} & \multicolumn{1}{l|}{51,20} & \multicolumn{1}{l|}{30,59} & \multicolumn{1}{l|}{35,50} \\ \cline{2-8} 
\multicolumn{1}{l|}{\textbf{ta01}} & \multicolumn{1}{l|}{1231} & \multicolumn{1}{l|}{1375} & \multicolumn{1}{l|}{1413} & \multicolumn{1}{l|}{1392} & \multicolumn{1}{l|}{13,70} & \multicolumn{1}{l|}{8,65} & \multicolumn{1}{l|}{2,50} \\ \cline{2-8} 
\multicolumn{1}{l|}{\textbf{ta10}} & \multicolumn{1}{l|}{1241} & \multicolumn{1}{l|}{1476} & \multicolumn{1}{l|}{1510} & \multicolumn{1}{l|}{1488,4} & \multicolumn{1}{l|}{11,91} & \multicolumn{1}{l|}{9,50} & \multicolumn{1}{l|}{3,68} \\ \cline{2-8} 
\multicolumn{1}{l|}{\textbf{ta40}} & \multicolumn{1}{l|}{1651} & \multicolumn{1}{l|}{2125} & \multicolumn{1}{l|}{2201} & \multicolumn{1}{l|}{2153} & \multicolumn{1}{l|}{26,14} & \multicolumn{1}{l|}{24,00} & \multicolumn{1}{l|}{10,71} \\ \cline{2-8} 
\multicolumn{1}{l|}{\textbf{ta20}} & \multicolumn{1}{l|}{5183} & \multicolumn{1}{l|}{6039} & \multicolumn{1}{l|}{6105} & \multicolumn{1}{l|}{6066,6} & \multicolumn{1}{l|}{24,82} & \multicolumn{1}{l|}{10,50} & \multicolumn{1}{l|}{9,42} \\ \cline{2-8} 
\end{tabular}
\end{table}

In this Appendix, it is shown the performance of the ACO for several other instances from \cite{Instances2020}. The parameters used during evaluation are present in Table~\ref{tab:param_perf} for $10$ executions each. The results are shown in Table~\ref{tab:extraresults}, where the algorithm suggests that it can achieve optimum or near optimum results for several instances.

\end{appendices}

\end{document}